\documentclass{Interspeech2024}

\interspeechcameraready

\title{Speakers Unembedded: Embedding-free Approach to Long-form Neural Diarization}

\name{Xiang}{Li}
\name{Vivek}{Govindan}
\name{Rohit}{Paturi}
\name{Sundararajan}{Srinivasan}

\address{
  AWS AI Labs
\email{\{xiangzai, vvekg, paturi, sundarsr\}@amazon.com}
}

\begin{document}

\maketitle

\begin{abstract}
    
End-to-end neural diarization (EEND) models offer significant improvements over traditional embedding-based Speaker Diarization (SD) approaches but falls short on generalizing to long-form audio with large number of speakers. EEND-vector-clustering method mitigates this by combining local EEND with global clustering of speaker embeddings from local windows, but this requires an additional speaker embedding framework alongside the EEND module. In this paper, we propose a novel framework applying EEND both locally and globally for long-form audio without separate speaker embeddings. This approach achieves significant relative DER reduction of 13\% and 10\% over the conventional 1-pass EEND on Callhome American English and RT03-CTS datasets respectively and marginal improvements over EEND-vector-clustering without the need for additional speaker embeddings. Furthermore, we discuss the computational complexity of our proposed framework and explore strategies for reducing processing times.
\end{abstract}
\section{Introduction}

Speaker diarization addresses the ``who spoken when" problem by partitioning an audio stream containing multiple speakers into homogeneous segments associated with each speaker. Conventional diarization systems \cite{anguera2012speaker, sell2014speaker, snyder2016deep, garcia2017speaker, sell2018diarization, wang2018speaker} typically consist of a cascade of several separate modules: voice activity detection to detect the speech frames, speaker embedding extraction to transform the speech segments into discriminative representations, and clustering to group speech regions by speaker identity. While effective for long-form audio with an arbitrary number of speakers, these cascaded multi-module approaches face challenges in handling overlapping speech and can suffer from error propagation across the modules.

To overcome the limitations of cascaded approaches, end-to-end neural diarization (EEND) was proposed in \cite{fujita2019end} which formulates speaker diarization as a frame-wise multi-label classification task with permutation invariant training \cite{yu2017permutation}. EEND can naturally handle overlapping speech by allowing multiple speakers to be active simultaneously and is also fully supervised compared to the unsupervised clustering component of the cascaded approach. However, despite its theoretical promise, EEND and its variants like EEND-SA \cite{fujita2019endsa}, EEND-EDA \cite{horiguchi2020end}, etc have struggled to generalize to larger numbers of speakers and arbitrarily long conversations.

In order to apply EEND models to longer audios and larger number of speakers, recent works \cite{kinoshita2021integrating, kinoshita2021advances} have proposed hybrid frameworks that integrate EEND with conventional clustering-based approaches. These methods leverage the strong diarization capability of EEND for speaker labeling over short local windows while performing global clustering on speaker embeddings computed across the local windows. This hybrid approach can handle both overlapping speech locally and long conversations with an arbitrary number of speakers globally. Most of the recent EEND improvements have focused on integrating additional embedding \cite{kinoshita2021integrating, kinoshita2021advances, zeghidour2021dive, yu2022auxiliary, fung2023robust, wang2023end, plaquet23_interspeech} or attractor modules \cite{horiguchi2021towards, rybicka2022end, fujita2023neural, hao2023end, chen2023attention, samarakoon2023transformer, landini2023diaper}, which requires specialized model architectures, loss functions and data requirements. Moreover, in some real-world scenarios, creating and storing speaker embeddings may need to be avoided where possible due to privacy considerations \cite{teixeira2023privacyoriented}.

In this paper, we propose a novel embedding-free approach that doesn't require any speaker embeddings and can still leverage the benefits of EEND and scale it to long-form audios with arbitrary number of speakers. We achieve this by utilizing a vanilla EEND model for both local diarization within the short local windows as well as global diarization across local windows, hence named local-global EEND. The proposed method consists of three steps: local EEND, global EEND, and clustering. In the local step, long audio is split into fixed-size windows, and EEND performs diarization within each window. The global step solves the inter-window label permutation by re-applying EEND to chunks formed by pairing speaker chunks across local windows. This generates pairwise speaker scores which are used to build an affinity matrix for the final clustering and global speaker labeling, without requiring any speaker embeddings.

The rest of the paper details the local-global EEND approach, experimental setup, results compared to the baselines, and a discussion on potential computation improvements for the global step.

\section{Local-global EEND}

\begin{figure*}[t]
  \centering
  \includegraphics[width=\linewidth]{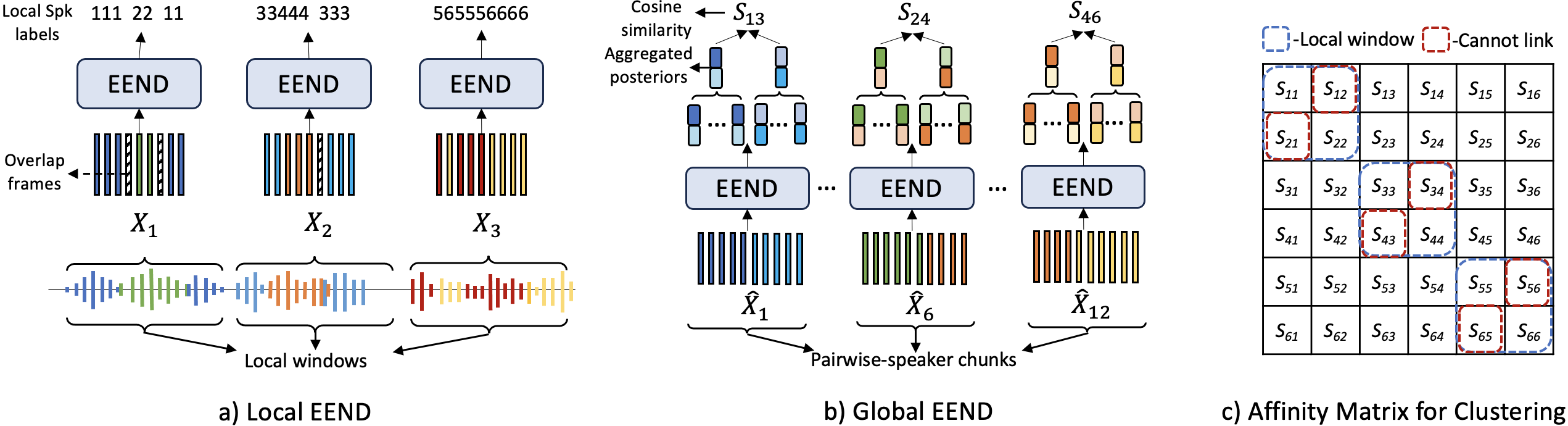}
  \caption{Local-global EEND framework. This assumes 3 local windows with 2-speaker local EEND, i.e $W$=3, $S_{local}$=2 resulting in $C$=12 pairwise-speaker chunks for global EEND.}
  \label{fig:local_global_eend}
\end{figure*}

\subsection{Local EEND}
Figure 1 shows the schematic diagram of the proposed embedding-free approach which can be divided into local EEND, global EEND and clustering steps.

The input audio is first split into $W$ windows with a fixed window length. In each window $i$, frame-level acoustic features are extracted, denoted as $\textbf{X}_i = \{\textbf{x}_{i,t}\}_{t=1}^{T}, \textbf{x}_{i,t}\in\mathbb{R}^F$ where $t$ is the frame index, $T$ is the total number of frames in a window and $F$ is the feature dimension of Mel-filterbank features in this work. Speaker label $\textbf{y}_{i,t}=\{y_{i,t,s}\}_{s=1}^{S_{local}}$ denotes speech activities for $S_{local}$ speakers at frame $t$ within window $i$ and is defined as
\begin{align}
    y_{i,t,s}=\begin{cases}
    0&\text{(Speaker $s$ is inactive at $t$)}\\
    1&\text{(Speaker $s$ is active at $t$)}\\
    \end{cases}
\end{align}
The local EEND estimates frame-wise posteriors $P(y_{i,t,s}|{\textbf{X}_i})$ in each window using a vanilla EEND model. These posteriors are  binarized using a threshold $Th_{local}$ and median filtered \cite{fujita2019endsa} to obtain the local speaker labels $\textbf{y}_{i,t}$. 

\subsection{Global EEND}
In order to perform global SD, the global EEND step computes the speaker similarities across the local windows using the same EEND model. In order to compute these, the overlapping speaker frames within each local window are first filtered out 
and the remaining frames of each speaker in a window are paired with the frames of speakers in subsequent windows, resulting in new chunks $\{\hat{\textbf{X}}_{i}\}_{i=1}^{C}$
\begin{equation} \tag{1}
\hat{\textbf{X}}_{i} = concat(\textbf{x}_{j,t} | t=\{m_{j,s}\}_{s=1}^{M}, \textbf{x}_{k,t} | t=\{n_{k,s}\}_{s=1}^{N})
\end{equation}
where $\textbf{x}_{j,t}$ and $\textbf{x}_{k,t}$ represent frame-level acoustic features of window $j$ and  window $k$ $(j\not=k)$, respectively. $\{m_{j,s}\}_{s=1}^{M}$ represents the $M$ frame indices corresponding to speaker $m$ in window $j$ and $\{n_{k,s}\}_{s=1}^{N}$ represents the $N$ frame indices corresponding to speaker $n$ in window $k$. $C$ is the total number of pairwise-speaker chunks processed by global EEND, where
\begin{equation} \tag{2}
C \leq W \times (W-1)/2 \times {S_{local}}^2
\end{equation}
In the case where a speaker has limited or no non-overlapping frames, we leverage the overlapping frames similarly to the EEND-vector clustering approach.

EEND is applied to $\hat{\textbf{X}}_{i}$ to generate inter-window frame-level speaker posteriors
\begin{equation} \tag{3}
[\textbf{z}_1, ...,\textbf{z}_{M}, \textbf{z}_{M+1}, ..., \textbf{z}_{M+N} ] = EEND(\hat{\textbf{X}}_{i})
\end{equation}
where $\{\textbf{z}_t\}_{t=1}^{M+N}, \textbf{z}_t \in\mathbb{R}^{S_{local}}$ are the inter-window frame-level posteriors. $[\textbf{z}_1, ...,\textbf{z}_{M}]$ and $[\textbf{z}_{M+1}, ..., \textbf{z}_{M+N}]$ are the posteriors corresponding to the $M$ frames of speaker $m$ and $N$ frames of speaker $n$ respectively. This process is repeated on every speaker pair across local windows as shown in Figure 1. 

\subsection{Embedding-free clustering}

The frame-wise posteriors $\{\textbf{z}_t\}_{t=1}^{M+N}$ are aggregated on frames belonging to the same speaker, resulting in speaker-level posteriors $\overline{\textbf{z}}_m$ and $\overline{\textbf{z}}_n$. Pairwise-speaker similarity $S_{mn}$ is then calculated as
\begin{equation} \tag{4}
\overline{\textbf{z}}_m = mean([\textbf{z}_1, ...,\textbf{z}_{M}])
\end{equation}
\begin{equation} \tag{5}
\overline{\textbf{z}}_n = mean([\textbf{z}_{M+1}, ..., \textbf{z}_{N}])
\end{equation}
\begin{equation} \tag{6}
S_{mn} = cosine\_simalirity(\overline{\textbf{z}}_m, \overline{\textbf{z}}_n)
\end{equation}
Each $S_{mn}$ is an entry of the affinity matrix $\textbf{S}\in\mathbb{R}^{S_{Global}\times S_{Global}}$ which will be used for the final clustering, where $S_{Global}$ is the sum of number of speakers detected in each local window with the local EEND, upper bound of which will be $W \times S_{local}$.

In order to enhance the clustering performance as well as to save on additional computations, we incorporate cannot-link constraints among different speakers identified within the same local window obtained in the local EEND step. This constraint is enforced by assigning a speaker similarity of 0 between local speaker pairs. Spectral clustering is then employed to group the speaker frames into $D$ speaker sets using the max eigengap heuristic similar to \cite{kinoshita2021advances, wang2018speaker}.

\begin{table*}
  \caption{Effect of sequence concatenation on CALLHOME2. $EEND_{vanilla}$ is the EEND model adapted on unmodified utterances. $EEND_{concat}$ is the EEND model adapted on speaker concatenated sequences described in \ref{subsec:seq concat}. $EEND_{vanilla}+EEND_{concat}$ uses $EEND_{vanilla}$ for the local step and $EEND_{concat}$ for the global step of our local-global approach.}
  \label{tab:seq_concate}
  
  \centering
  \begin{tabular}{cccccccc}
    \toprule
    \multirow{2}*{System} & \multirow{2}*{Model} & \multicolumn{6}{c}{\# of speakers in a session} \\
    \cline{3-7} & \\[-1.0em]
    & & 2 & 3 & 4 & 5 & 6 & all \\
    \midrule
    \multirow{2}*{1-pass EEND} & $EEND_{vanilla}$ & 7.53 & 14.91 & - & - & - & - \\
    & $EEND_{concat}$ & 7.36 & 17.74 & - & - & - & - \\
    \midrule
    \multirow{3}*{Local-global EEND} & $EEND_{vanilla}$ & 7.99 & 12.21 & 16.39 & 17.10 & 26.12 & 12.48 \\
    & $EEND_{concat}$ & \textbf{7.29} & 11.85 & 17.83 & \textbf{15.76} & \textbf{22.38} & \textbf{12.16} \\
    & $EEND_{vanilla}+EEND_{concat}$ & 7.66 & \textbf{11.67} & \textbf{16.03} & 17.56 & 23.71 & 12.45 \\
    
    \bottomrule
  \end{tabular}
  
\end{table*}

\section{Experiments}

In this section, we go over the datasets used, model architecture, settings and techniques followed for efficient inference. 

\subsection{Data and metrics}

For training the EEND model, we generate simulated mixtures by mixing Switchboard-2 (Phase I \& II \& III), Switchboard Cellular (Part 1 \& 2), and the NIST Speaker Recognition Evaluation (2004 \& 2005 \& 2006 \& 2008) with MUSAN corpus \cite{snyder2015musan}, following the data generation procedure in \cite{fujita2019end}. Mixtures with up to 3 speakers were created, with $\beta=[2, 2, 9]$ for mixture with 1, 2 and 3 speakers, respectively. 

For model adaptation and evaluation, real telephone conversation dataset CALLHOME \cite{przybocki2000martin}, i.e., NIST SRE2000 (LDC2001S97, Disk-8) is used. It is widely used as the benchmark for existing EEND-based approaches. The CALLHOME dataset contains 500 sessions, each with 2 to 6 speakers. There are mostly two dominant speakers in each conversation. We split the data into two subsets according to \cite{horiguchi2020end} for adaptation (CALLHOME1) and evaluation (CALLHOME2). 

As local-global EEND framework is designed for dealing with long conversations, to showcase the effectiveness of this framework, evaluations on other benchmarks with longer audios are reported as well, such as CALLHOME American English (CHAE) \cite{canavan1997nist} and RT03-CTS \cite{fiscus2007nist} which have an average duration of 30 and 10 minutes respectively. We use the official eval splits for evaluation on these datasets.

For evaluation metrics, we use the standard Diarization Error Rate (DER) \cite{nist2009evaluation} with a collar tolerance of 250ms and included the overlapping speech segments while scoring.

\subsection{EEND model settings}

To ensure a fair comparison with the existing hybrid baseline, we adopt the front-end configuration from EEND-vector-clustering \cite{kinoshita2021advances}. This involves the extraction of 23-dimensional log-Mel-filterbank features, utilizing a frame length of 25ms and a frame shift of 10ms. The window size $T$ is set at 300 (=30s) for both training and adaptation. The EEND architecture consists of 6 stacked self-attention-based Transformer layers, featuring eight attention heads and a hidden size of 256. This aligns with the configuration employed in \cite{kinoshita2021advances}. In each window, the EEND model estimates the posteriors for up to 3 speakers.

During both training and adaptation, we employ the Adam optimizer \cite{kingma2014adam} alongside the Noam scheduler \cite{vaswani2017attention}, incorporating 150,000 warm-up steps for training. For adaptation, a fixed learning rate of $1\times10^{-5}$ is utilized. Both training and adaptation phases span 100 epochs.

\subsection{Sequence concatenation during adaptation} \label{subsec:seq concat}

In the global EEND step, we generate the chunk-level input by concatenating frame-level acoustic features between every pair of speakers across local windows. This process results in a new pairwise-speaker sequence that has not been encountered in either the training or adaptation data. To enhance EEND's generalization to this new input format, we incorporate this data generation procedure during adaptation. First, the frame-level acoustic features from the same speaker in each utterance are aggregated into several blocks after discarding the overlapping speech frames. Subsequently, every two blocks are concatenated to generate a new input utterance. We reformat the data using this technique for half of the samples in each batch. 



\subsection{Efficiency improvement for inference}

Real Time Factor (RTF) is a criteria used to measure the efficiency of SD systems. It is calculated by dividing the time taken by the SD system by the total duration of the spoken audio. 

In the global step during inference, each speaker within a local window is paired with every speaker in subsequent local windows, resulting in a computational complexity of $O(n^2)$. As illustrated in Figure 1, if there are 3 local windows, each containing 2 speakers, there will be 12 inference calls in the global EEND step. 
To enhance GPU efficiency and reduce RTF, we propose batching multiple inference requests together. Additionally, we explore different number of random frames ($N=128, 64, 32, 16$) to minimize the number of frames required for each speaker during global EEND inference, thereby reducing computational load.



\section{Results}

In this section, we present the outcomes of our experiments, beginning with an evaluation of the impact of sequence concatenation in the adaptation process. Subsequently, we compare the proposed approach with existing baselines, utilizing both ora- cle and estimated speaker counts during clustering. Our analysis extends to additional benchmarks, such as CALLHOME American English (CHAE) and RTCTS. Finally, we delve into an examination of the efficiency improvements in global EEND inference.

\subsection{Effect of sequence concatenation}

We explored two types of input data formats during adaptation, resulting in $EEND_{vanilla}$ and $EEND_{concat}$. The former denotes the model adapted with original input sequences typically used in EEND model adaptation, while the latter is adapted with a combination of original sequences and concatenated sequences, as described in Section \ref{subsec:seq concat}. Table 1 presents a comparison between the local-global EEND and 1-pass EEND, utilizing the different adaptation techniques. We used oracle speaker information during global clustering and selected a binarization threshold between 0.3 to 0.7 that produced the best DER on the validation set for both 1-pass and local EEND. We only evaluated 1-pass EEND on 2,3 speaker sessions since it was trained to only detect a maximum of 3 speakers.

Local-global EEND outperforms 1-pass EEND on $EEND_{vanilla}$ by 18\% in the 3-speaker session but shows a marginal performance degradation in the 2-speaker session whereas $EEND_{concat}$ outperforms the best 1-pass EEND by 3\% and 21\% in 2-speaker and 3-speaker sessions, respectively.

For 1-pass EEND, $EEND_{concat}$ only marginally benefits the 2-speaker session but not the 3-speaker one, as expected due to the data mismatch between adaptation and evaluation, where concatenation during adaptation only occurs on pairwise speakers. When evaluating more speakers for local-global EEND, $EEND_{concat}$ consistently performs better then $EEND_{vanilla}$, except for the 4-speaker session.

In order to exactly match the local and global input conditions, we also attempted to apply $EEND_{vanilla}$ in local EEND and $EEND_{concat}$ in global EEND. This further improved performance in 3 and 4 speaker sessions but $EEND_{concat}$ for both steps achieves the best overall DER across all speaker sessions.

\begin{table}
  \caption{Comparison with other baselines with oracle number of speakers on CALLHOME2. The best
scores are \textbf{bolded}.}
  \label{tab:seq_concate}
  \centering
  \addtolength{\tabcolsep}{-3pt}
  \scalebox{0.8}{%
  \begin{tabular}{ccccccc}
    \toprule
    \multirow{2}*{System} & \multicolumn{5}{c}{\# of speakers in a session} \\
    \cline{2-6} & \\[-1.0em]
    & 2 & 3 & 4 & 5 & 6 & all \\
    \midrule
    x-vector-clustering \cite{horiguchi2020end} & 8.93 & 19.01 & 24.48 & 32.14 & 34.95 & 18.98 \\
    EDA-EEND \cite{horiguchi2020end} & 8.35 & 13.20 & 21.71 & 33.00 & 41.07 & 15.43 \\
    EEND-vector-clust. (T=30s) \cite{kinoshita2021advances} & 8.08 & \textbf{11.27} & \textbf{15.01} & 23.14 & 26.56 & 12.22 \\
    
    \midrule
    Local-global EEND & \textbf{7.29} & 11.85 & 17.83 & \textbf{15.76} & \textbf{22.38} & \textbf{12.16}\\
    
    \bottomrule
  \end{tabular}}
  
\end{table}

\begin{table}[t]
  \caption{Comparison with other baselines with estimated number of speakers on CALLHOME2. The best
scores are \textbf{bolded} and the second best are \underline{underlined}.}
  \label{tab:seq_concate}
  \centering
  \addtolength{\tabcolsep}{-3pt}
  \scalebox{0.8}{%
  \begin{tabular}{ccccccc}
    \toprule
    \multirow{2}*{System} & \multicolumn{5}{c}{\# of speakers in a session} \\
    \cline{2-6}& \\[-1.0em]
    & 2 & 3 & 4 & 5 & 6 & all \\
    \midrule
    x-vector-clustering \cite{horiguchi2020end} & 15.54 & 18.01 & 22.68 & 31.40 & 34.27 & 19.43 \\
    EDA-EEND \cite{horiguchi2020end} & 8.50 & 13.24 & 21.46 & 33.16 & 40.29 & 15.29 \\
    EEND-vector-clust. (T=30s) \cite{kinoshita2021advances} & 7.96 & \underline{11.93} & \underline{16.38} & \underline{21.21} & 23.10 & 12.49 \\
    EEND-EDA-local-global \cite{horiguchi2021towards} & \textbf{7.11} & \textbf{11.88} & \textbf{14.37} & 25.95 & \textbf{21.95} & \textbf{11.84} \\
    \midrule
    Local-global EEND & \underline{7.51} & 12.20 & 17.88 & \textbf{16.01} & \underline{22.35} & \underline{12.20} \\
    
    \bottomrule
  \end{tabular}}
  
\end{table}

\subsection{Comparison with other baselines}

The performance of local-global EEND is compared with other baselines in Table 2 and Table 3, with oracle and estimated numbers of speakers respectively. Results for EEND-vector-clustering with a window size of 30s are extracted from their paper, and the outcomes for local-global EEND with $EEND_{concat}$ are reported. 

In the case of oracle numbers of speakers, local-global EEND outperforms EEND-vector-clustering \cite{kinoshita2021advances} in all sessions except for 3 and 4-speaker sessions, with only a marginal degradation in the 3-speaker session. Particularly noteworthy is the substantial improvement of 32\% and 15.7\% in 5- and 6-speaker sessions, respectively. The decline in the 4-speaker session may be attributed to the sub-optimal window size, a phenomenon observed in the EEND-vector-clustering paper as well, suggesting that different window sizes may result in significant performance variations.

Table 3 presents results with estimated numbers of speakers, introducing another strong baseline, EEND-EDA-local-global \cite{horiguchi2021towards}. Compared to EEND-vector-clustering, local-global EEND achieves superior performance across nearly all sessions, with similar exceptions in the 4-speaker session. In comparison to EEND-EDA-local-global, local-global EEND exhibits slightly inferior performance in the $\{2, 3, 6\}$-speaker sessions but significantly outperforms in the 5-speaker session. This discrepancy could stem from differences in training data volume where the local-global EEND is trained with only up to 3-speaker mixtures, whereas EEND-EDA-local-global is trained on a larger training data including 100,000 additional 4-speaker mixtures. Consequently, EEND-EDA-local-global achieves the best DER in the 4-speaker session, aligning with this matched training scenario.

\begin{table}[t]
  \caption{DER (\%) on long-form audio datasets ($Th_{local}$=0.5).}
  \label{tab:seq_concate}
  \centering
  \scalebox{0.9}{%
  \begin{tabular}{cccc}
    \toprule
    \multirow{2}*{System} & \multicolumn{3}{c}{Dataset}  \\
    \cline{2-4} & \\[-2.0ex]
    & CHAE test & CHAE 109 & RTCTS test\\
    \midrule
    1-pass EEND & 5.95 & 7.25 & 5.69\\
    Local-global EEND & 5.20 & 7.02 & 5.14\\
  
    \bottomrule
  \end{tabular}}
  
\end{table}

\subsection{Performance on other benchmarks}

To provide a comprehensive evaluation of the local-global EEND system, we extend our analysis to other well-established diarization benchmarks, namely CALLHOME American English (CHAE) and RTCTS. These datasets feature longer-duration audios, offering insights into the efficacy of local-global diarization systems. Binarization is carried out using a fixed threshold ($Th_{local}$=0.5) to obtain diarization results, and clustering is performed using the estimated number of speakers.

As depicted in Table 4, local-global EEND demonstrates superior performance compared to 1-pass EEND across various datasets. Specifically, it outperforms 1-pass EEND by 12.7\%, 9.7\% and 3\%, on the CHAE test set, RTCTS test set and CHAE 109, respectively.

\subsection{Inference efficiency improvements}
Figure 2 shows the results on CHAE test set with different strategies to improve inference efficiency. All the experiments are performed on NVIDIA A10G GPU on AWS cloud (G5-2xLarge). Moving from sequential inference to batching (batching 500 chunks), the RTF is reduced by 50\%. Further RTF reduction is gained from reducing computation by selection a subset of frames (N=128, 64, 32, 16) randomly for each speaker in global EEND. A subset of 64 frames can produce a desirable RTF reduction by nearly 70\% with no impact on DER. Regarding the computational cost versus the input audio length, the local-global EEND produces the RTF of $\{7.3 e^{-3}, 1.5 e^{-2}, 2.2 e^{-2}, 5.0 e^{-2}\}$ for the audio length of $\{5, 10, 15, 30\} mins$, respectively. 

\begin{figure}[t]
  \vspace{1.0\baselineskip}
  \centering
  \includegraphics[width=\linewidth]{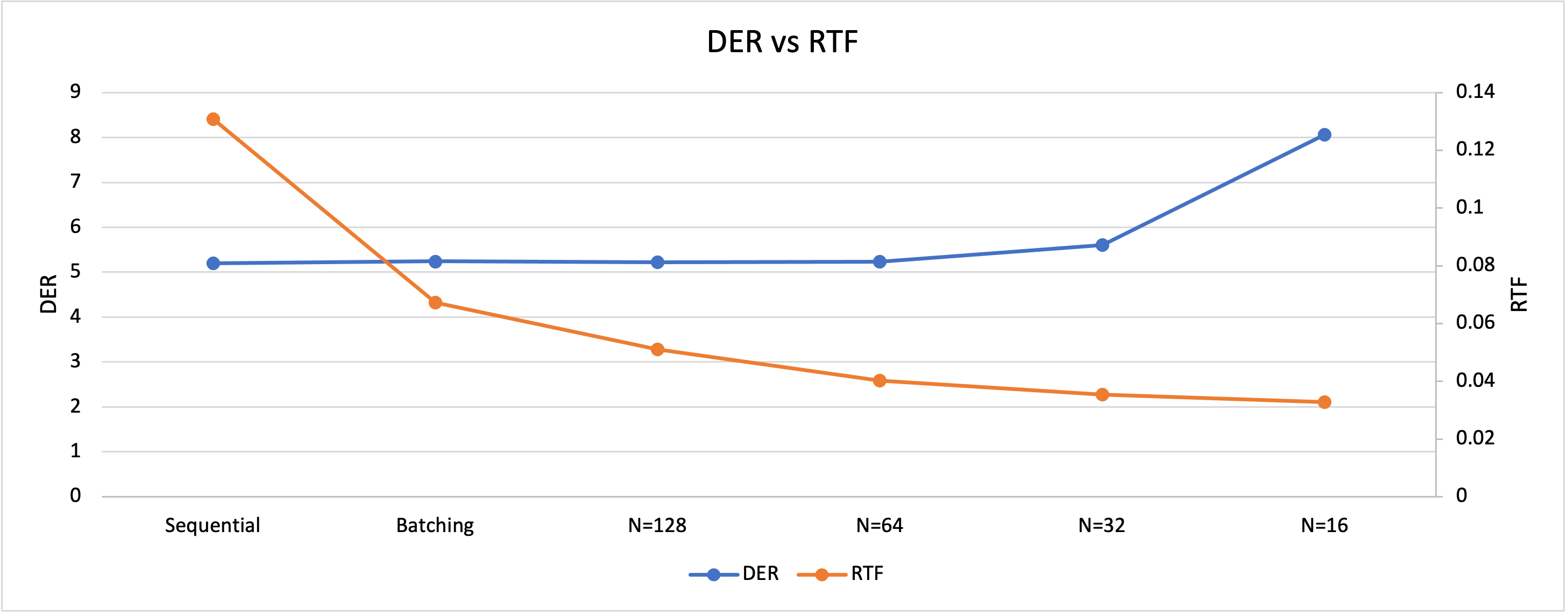}
  \caption{RTF vs DER with different strategies on efficiency improvement, including batching the inferences and minimizing the number of frames required for each speaker. N indicates a subset of N random frames.}
  \label{fig:RTF_vs_DER}
\end{figure}



\section{Conclusion}

This paper introduces a novel embedding-free diarization methodology that employs EEND in both local and global steps. The global clustering is accomplished without the need for speaker embeddings, utilizing EEND on concatenated pairwise speaker features across local windows to derive the pairwise speaker similarities. This approach achieves significant relative DER reduction of 13\% and 10\% over the conventional 1-pass EEND on CHAE and RT03-CTS datasets respectively and even offers a marginal 3\% relative DER reduction over EEND-vector-clustering without the need for additional speaker embeddings or loss functions. The paper also includes a discussion on the computational complexity of the global EEND step and explores strategies for reducing the processing times. By batching multiple chunk-level inferences and minimizing the number of frames required for each speaker, the RTF can be reduced by nearly 70\% without the impact on diarization performance.


\bibliographystyle{IEEEtran}
\bibliography{mybib}

\end{document}